\documentclass[amsmath,amssymb,amsfonts,twocolumn]{revtex4-2}
\usepackage{graphicx}
\usepackage{csquotes}
\usepackage{float}
\usepackage{braket,bigints,stackengine,scalerel}
\usepackage{stmaryrd,mathtools}
\def \beq {\begin{equation}}
\def \eeq {\end{equation}}
\def \tr {\rm Tr}
\def \qs {\rm Q_ S}
\def \qt {\rm Q_T}
\begin{document}
\title{Physiological search for quantum biological effects based on the Wigner-Yanase connection between coherence and uncertainty} 
\author{I. K. Kominis}
\affiliation{Department of Physics and Institute for Theoretical and Computational Physics, University of Crete, 71003 Heraklion, Greece}
\begin{abstract}
A fundamental concept of quantum physics, the Wigner Yanase information, is here used as a measure of quantum coherence in spin-dependent radical-pair reactions pertaining to biological magnetic sensing. This measure is connected to the uncertainty of the reaction yields, and further, to the statistics of a cellular receptor-ligand system used to biochemically convey magnetic-field changes. Measurable physiological quantities, such as the number of receptors and fluctuations in ligand concentration, are shown to reflect the introduced Wigner-Yanase measure of singlet-triplet coherence. We arrive at a quantum-biological uncertainty relation, connecting the product of a biological resource and a biological figure of merit with the Wigner-Yanase coherence. Our approach can serve a general search for quantum-coherent effects within cellular environments.
\end{abstract}
\maketitle

\section{Introduction}
Quantum biology promises to push the quantum-to-classical transition barrier \cite{Plenio}, and to some extent include into the quantum world the more complex setting of biological systems. Spin is an archetypal quantum system, often weakly coupled to decohering environments \cite{Kanai}, hence retrospectively it is not surprising that spin-chemical effects in biomolecular reactions \cite{Ritz,Hore1,Hore2} have become a major paradigm for quantum biology \cite{PRE2009,PRE2011,CPL2012,NJP2013,PRE2014,KominisReview,PRE2017,PRA2017,PRR2020}. Excitation transfer in photosynthesis \cite{Guzik,P,Zigmantas1,Duan,Scholes,Zigmantas2,Cao,Miller} and quantum spectroscopic effects in olfaction \cite{Turin,Skoulakis,Brookes} have also been studied in the context of quantum biology, with more possibilities being contemplated \cite{Fisher,Mazzoccoli}. 

Apart from a few specific examples supporting the premise of quantum biology, a lingering question is whether nature has invented quantum technology and widely applied it, or such paradigms are just a few outliers in a generally classical biological world. As biological processes are virtually limitless, a case-by-case study of the detailed underlying physical mechanisms at temporal and spatial scales conducive to quantum-coherent effects is rather challenging. So far, this is the only available option.

We here introduce a systematic search methodology based on an intimate connection between quantum coherence and quantum uncertainty. If quantum-coherent dynamics are at the core of a biological process, then such process could have evolved towards an economic design wherein statistical effects in layers beyond the quantum-coherent core have adapted to the core's statistical uncertainties. This is not an unrealistic assumption, as for example, the sensitivity of the dark-adapted human visual system is limited by the photon statistics of the stimulus light \cite{Bialek}. To make the connection between coherence and uncertainty we use a coherence measure derived from the Wigner-Yanase information \cite{WY}. 

The benefit of this approach is that there could be numerous biological processes, particularly within cellular environments, where quantum coherent oscillations might not be immediately accessible, as they are in two-dimensional electronic spectroscopy of light-harvesting systems \cite{Mukamel,Ogilvie}, or in spin-chemical reactions \cite{Lambert}, both studied outside of the cellular environment. We illustrate our findings with the radical-pair mechanism of biological magnetic sensing, and a cellular receptor-ligand system working as a biochemical transducer of the quantum-coherent process. The general scope of this approach becomes apparent by the final result, which connects a quantifier of coherence with measurable biochemical quantities, without any reference to the details of the coherent process. Finally, we arrive at a quantum-biological uncertainty relation. It connects the product of a biological resource (number of receptors) and a figure of merit (magnetic sensitivity) with the Wigner-Yanase singlet-triplet coherence.
\section{This work in the context of previous work on quantum biology}
\begin{figure*}[t]
\begin{center}
\includegraphics[width=16 cm]{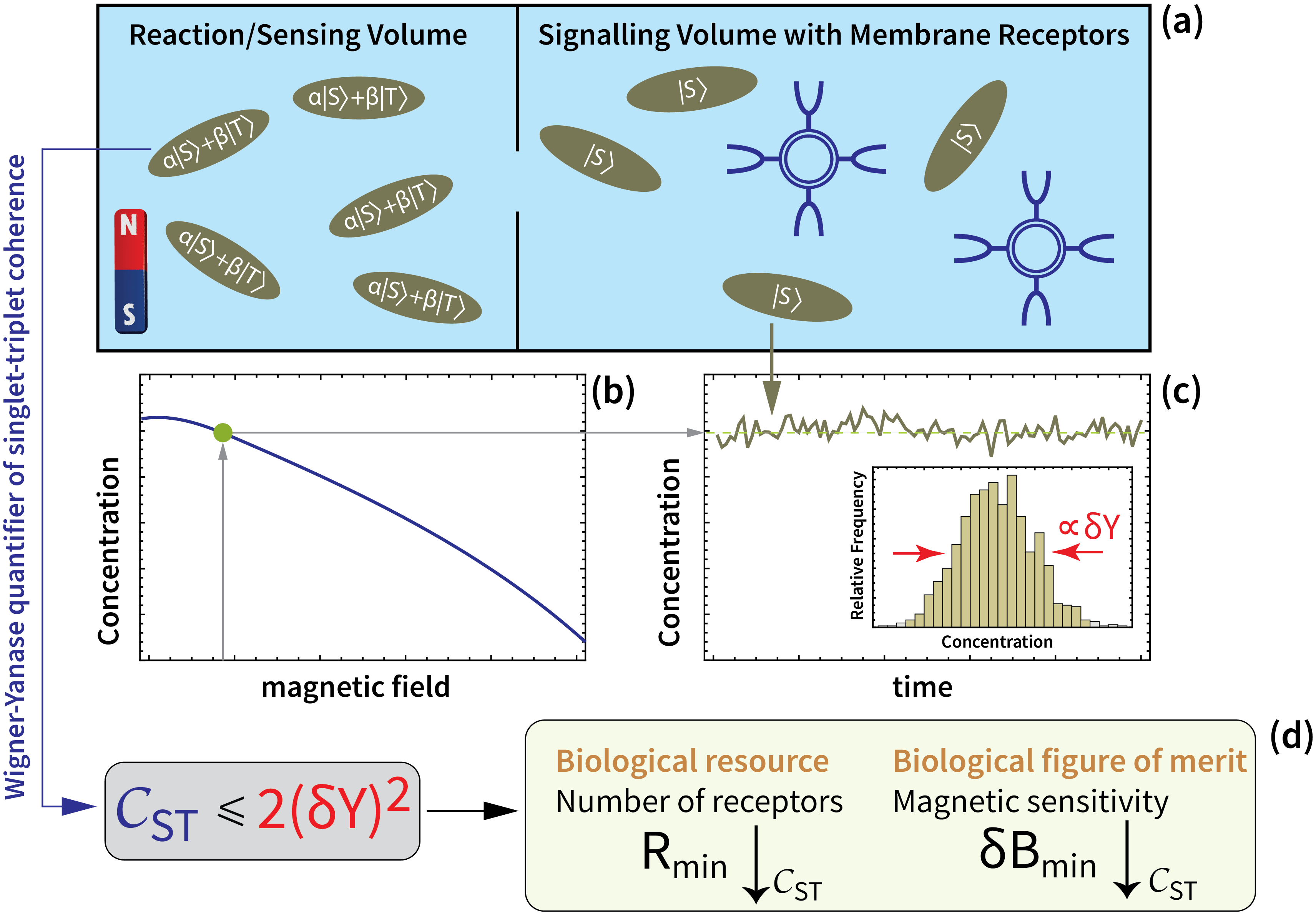}
\caption{Basic ideas of this work. (a) A biochemical spin-dependent reaction (radical-pair reaction) takes place in the reaction/sensing volume. Biomolecular spin states evolve in time, being in a (partially) coherent singlet-triplet superposition, here shown for simplicity as a pure state. The spin-state evolution is also affected by the ambient magnetic field, hence this reaction acts as a biochemical magnetometer. The reaction products are found into either singlet or triplet states, and the singlet products are assumed to enter a signaling volume, where they are detected by binding onto membrane-bound receptors. The signaling model was developed in \cite{Weaver}. (b) The concentration of the singlet reaction products depends on the magnetic field, with the dependence given by the radical-pair reaction dynamics. (c) The concentration of singlet products as a function of time at some operating point (some particular magnetic field), showing the fluctuations around a given (magnetic-field-dependent) value. The relevant distribution shown in the inset has a width proportional to $\delta Y$, the quantum uncertainty in the singlet reaction yield. In this work, we use a coherence measure based on the Wigner-Yanase information and connect the coherence of the reactants (the biomolecular spin states in the reaction volume) with $\delta Y$, the fluctuations in the reaction products. Our main finding is a formal bound for the reaction-averaged Wigner-Yanase coherence, ${\cal C}_{\rm ST}$, shown to be bounded above by twice the quantum variance of the singlet reaction yield, $2(\delta Y)^2$. Using the signaling model of \cite{Weaver}, we then connect the minimum number of receptors $R_{\rm min}$, which is a biological resource, and the magnetic sensitivity $\delta B_{\rm min}$, which is a biological figure of merit, with the reaction's coherence ${\cal C}_{\rm ST}$, showing that both $R_{\rm min}$ and $\delta B_{\rm min}$ decrease with increasing coherence.}
\label{intro}
\end{center}
\end{figure*}
As this work connects quantum coherence measures with biochemical reactions and biochemical signaling processes, we first make a few introductory remarks aimed at making these connections clear to the general reader, putting this work in the context of previous work. 

Radical-pair reactions form the central biochemical system studied by spin chemistry, a field addressing the effects of spin (electronic or nuclear) on chemical reactions \cite{Steiner}. Spin chemistry is an active field since the 1960's, however it was relatively recently that such spin-dependent biochemical reactions were treated as an open quantum system with the tools of modern quantum information science \cite{PRE2009}. In particular, it was shown that a number of quantum measurement processes take place during such reactions, and the fundamental decoherence processes were unraveled \cite{KominisReview}. 

In more detail, it was  understood for a long time that coherent oscillations between an electronic singlet and an electronic triplet state are central in driving reaction dynamics, and in fact leading to magnetic-sensitive reaction yields \cite{Timmel1998}. On this premise was founded the early idea of Schulten that such reactions underlie the workings of the avian magnetic compass \cite{Schulten}. However, the rich quantum-dynamic aspects of these reactions remained elusive until it was shown \cite{PRE2009} that the vibrational states of the biomolecules undergoing spin-selective reactions realize a continuous quantum measurement of the radical-pair's spin state. Several works \cite{Cai2011,Kais,Briegel,Shao,Li2017,Dasari} addressed quantum effects in these reactions, applying the sophisticated tools of quantum metrology and quantum information in a complex biophysical setting. Among the recent works, a promising approach was to simulate such reactions in a quantum computer \cite{IBM}. Radical-pair reactions are currently a major paradigm for quantum biology, since quantum processes at the molecular scale are seen to propagate even to the behaviour of the organism at the macroscopic scale, as for example in the case of the compass. Moreover, this paradigm has made the general premise of quantum biology realistic and worthwhile exploring further, since we are gradually learning how nature found intricate ways to work right at the boundary of the quantum-to-classical transition \cite{qct}, designing processes that are not maximally quantum-coherent yet not classical either.

Quantum coherence in radical-pair reactions and its potential role as a useful ''resource" has been studied in several works. An empirical (not formal) attempt to quantify singlet-triplet coherence was made in \cite{PRE2011}, while Plenio and coworkers formally addressed the global coherence of the radical-pair spin states \cite{Cai}. Since the singlet-triplet basis is central in radical-pairs due to the two recombination channels leading to two distinct reaction products, the singlet and the triplet, the quantification coherence in this particular basis is important. A formal a quantifier for singlet-triplet (S-T) coherence was introduced only recently by this author based on quantum relative entropy \cite{PRR2020}, further demonstrating S-T coherence is indeed a useful resource for the compass working of such reactions. Without such an explicit quantifier at their disposal, previous works \cite{HoreFD} attempted to understand the \enquote{quantumness} of radical-pair reactions using semi-classical approximations to the reaction dynamics and comparing with full quantum simulations.

In this work we introduce yet another formal quantifier of singlet-triplet coherence, this one based on the Wigner-Yananse information. The Wigner-Yanase information is a fundamental quantum-information-theoretic concept regarding a quantum system's density matrix $\rho$, and reflects uncertainty in   observables measured in the state $\rho$. Because of this connection to uncertainty, It is here shown that the singlet-triplet coherence measure based on the Wigner-Yanase information is rather useful, as it is connected with the uncertainty in the reaction-product yields. It thus allows the estimation of coherence, which might not be straightforward otherwise, by observing fluctuations in the concentration of particular biomolecules. 

Then, using a signaling model developed by Weaver and coworkers \cite{Weaver}, we here connect the statistics of the signaling system with the underlying coherence of the radical-pair reactions. In particular, the authors in \cite{Weaver} considered the binding of the radical-pair reaction products (the singlet products) to membrane receptors, since ligand-receptor binding is a ubiquitous signaling process in biology. We here introduce into the model of \cite{Weaver} one additional uncertainty, the quantum uncertainty $\delta Y$ of the singlet reaction yield. Assuming that the signaling system has been designed by nature to be quantum-limited, i.e. so that the quantum uncertainty $\delta Y$ is dominant, we then unravel connections between the reaction's coherence on the one hand, and on the other the minimum number of receptors required to achieve a given magnetic sensitivity. We thus connect a biological resource (number of receptors) and a biological figure of merit (magnetic sensitivity) with the underlying quantum coherence. All of the above are schematically summarised in Fig. \ref{intro}.

The quantum-limited design is a physically allowed assumption which, building on our formal bound connecting singlet-triplet coherence with fluctuations in reaction yields, leads to the aforementioned connections regarding the biological resource and figure of merit. It is moreover a realistic assumption, because, for example, the dark-adapted human visual system is known to be limited by the photon statistics of the stimulus light \cite{Bialek}. This is indeed a valid parallel, since the visual perception starts with a complex cascade of chemical reactions taking place in rod cells \cite{Rieke}, involving time-dependent concentrations of numerous signaling molecules, the end result of which is an action potential fired and transmitted to the brain. Visual perception rests on several such action potentials, and it is remarkable that the whole process is indeed so fine-tuned as to be sensitive to the photon statistics of the stimulus light. 

We here make a similar assumption regarding the design of the signaling process introduced by Weaver and coworkers. Namely, that the classical noise these authors consider is smaller than the quantum noise that necessarily accompanies the concentration changes of the radical-pair reaction products. Whether such an assumption is actually materialized in nature remains to be found. In case it is, however, we unravel the aforementioned connections between straightforward physiological observations and the underlying coherence of the reaction. Moreover, in a certain parameter regime we show that the biological resource (number of receptors) and the biological figure of merit (magnetic sensitivity) both gain an advantage with increasing coherence, but not arbitrarily. We show there is a constraint between those quantities, which looks like a quantum-biological uncertainty relation.

These results can help the search for quantum biological effects in cellular environments, since they lay out specific bounds and scaling laws of measurable quantities at the physiological level, and thus allow the search on how far a core quantum-coherent effect can propagate within cellular processes.
\section{Quantum fluctuations in radical-pair reaction yield}
We will first recapitulate radical-pair reaction dynamics, which form the paradigm quantum biological system used in this work. At the end of this discussion we will arrive at an expression for the quantum uncertainty of the reaction-product yield. Radical-pairs, pertinent to biological magnetic sensing, are interesting both in their own right as magnetic-sensitive chemical reactions \cite{Woodward1,Timmel,Woodward2,Hore1,Hore2}, and as a physical realization of the avian magnetic compass \cite{Ritz,compass1,compass2,compass3}. Radical-ion pairs are created by electron transfer from photoexcited donor-acceptor dyads DA (Fig. \ref{fig1}a). Initially the two unpaired electron spins are in the singlet state. Magnetic interactions captured by the spin hamiltonian ${\cal H}_B$ drive a coherent singlet-triplet (S-T) mixing $^{\rm S}{\rm D}^{\bullet +}{\rm A}^{\bullet -}\leftrightharpoons~^{\rm T}{\rm D}^{\bullet +}{\rm A}^{\bullet -}$. Dominant terms in ${\cal H}_B$ are the hyperfine coupling of the unpaired electron spins to the molecular nuclear spins, and the magnetic-field-dependent electronic Zeeman terms. S-T conversion involves a coherent spin motion of all spins involved, however, it is the coherence between the S and T subspace that is mostly of interest in reaction dynamics \cite{PRR2020}. This is because the end of the reaction through charge recombination takes place either from the electron spin singlet channel or from the electron spin triplet channel, leading correspondingly to two kinds of neutral reaction products, the singlet neutral products (DA) and the triplet neutral products (${\rm ^TDA}$). 

To quantify the recombination process, let $\rho_t$ describe the radical-pair spin state at time $t$. Within the time interval $dt$ there will be $dn_S$ singlet and $dn_T$ triplet neutral products, where $dn_x=k_xdt\tr\{\rho_t{\rm Q}_x\}$, with $x={\rm S,T}$. Here $\qs$ and $\qt$ are projectors to the radical-pair S and T subspace, while $k_S$ and $k_T$ are the corresponding recombination rates. For self-completeness we note that the projectors $\qs$ and $\qt$ are complete and orthogonal, i.e. $\qs+\qt=\mathsf{1}$ and $\qs\qt=0$, where $\mathsf{1}$ is the unit matrix of dimension $d=4d_{\rm nuc}$. The dimension $d$ of the radical-pair's density matrix is determined by the spin multiplicity of the two unpaired electrons (4) times the spin multiplicity, $d_{\rm nuc}=\Pi_{j=1}^{N_{\rm nuc}}(2I_j+1)$, of $N_{\rm nuc}$ magnetic nuclei having nuclear spin $I_j$, with $j=1,2,...N_{\rm nuc}$. Thus the two projectors split the radical-pair Hilbert space into an electronic singlet subspace of dimension $d_{\rm nuc}$ and an electronic triplet subspace of dimension $3d_{\rm nuc}$.

For equal recombination rates considered herein, $k_S=k_T\equiv k$, it is \cite{PRR2020} $\rho_t=e^{-kt}\tilde{\rho}_t$, where $\tilde{\rho}_t$ follows the trace-preserving evolution $d\tilde{\rho}_t/dt=-i[{\cal H}_B,\tilde{\rho}_t]-\kappa_{ST}(\qs\tilde{\rho}_t+\tilde{\rho}_t\qs-2\qs\tilde{\rho}_t\qs)$, with $\kappa_{ST}$ being the S-T dephasing rate. The dephasing term in $d\tilde{\rho}_t/dt$ can be considered to arise from an unobserved measurement in the S-T basis taking place during the radical-pair state evolution \cite{KominisReview}. In previous work we have shown that the intrinsic quantum dynamics of the reaction lead to $\kappa_{ST}=(k_S+k_T)/2$. However, here we take $\kappa_{ST}$ as a free parameter, in order to include additional sources of spin relaxation possibly leading to S-T dephasing. 

The S and T reaction yields are $Y_x=\int_0^\infty dn_x=\int_0^\infty dtke^{-kt}\tr\{\tilde{\rho}_tQ_x\}$, where $x={\rm S,T}$. Both yields depend on the magnetic field $B$ through $\tilde{\rho}_t$, the time evolution of which is influenced by the $B$-dependent hamiltonian ${\cal H}_B$. Since we will only be dealing with the singlet yield, we denote $Y\equiv Y_S$. The reaction yield $Y$ depends on the expectation value $\tr\{\tilde{\rho}_t\qs\}$, hence it is accompanied by a quantum uncertainty, $\delta Y$, stemming from the (time-dependent) uncertainty of $\qs$ at the radical-pair state $\tilde{\rho}_t$. In other words, by repeating the reaction several times, each time with $N_{\rm rp}$ radical-pairs entering the reaction, the number of singlet products will be distributed around the mean $YN_{\rm rp}$ with uncertainty $\delta Y\sqrt{N_{\rm rp}}$. 

To calculate $\delta Y$ we note that it is variances that add up, hence the variance $(\delta Y)^2$ consists of the time-dependent variances of $\qs$ in the radical-pair state $\rho_t$, denoted by $(\Delta \qs)_{\tilde{\rho}_t}^2$. The contribution of this variance depends on how far has the reaction proceeded, thus $(\Delta \qs)_{\tilde{\rho}_t}^2$ is weighted by the reaction term $ke^{-kt}$. Since $\qs$ is a projector, it is $\qs^2=\qs$, hence 
\begin{align}
(\Delta \qs)_{\tilde{\rho}_t}^2&=\tr\{\tilde{\rho}_t\qs^2\}-\tr\{\tilde{\rho}_t\qs\}^2\nonumber\\&=\tr\{\tilde{\rho}_t\qs\}(1-\tr\{\tilde{\rho}_t\qs\})
\end{align}
Since the singlet and triplet projectors form a complete set, $\qs+\qt=\mathsf{1}$, we can write $\delta Y$ in the S-T symmetric form 
\beq
\delta Y=\Big[\int_0^\infty dtke^{-kt}\tr\{\tilde{\rho}_t\qs\}\tr\{\tilde{\rho}_t\qt\}\Big]^{1/2}\label{dY}
\eeq
The uncertainty $\delta Y$ is generally of quantum origin, as it reflects the randomness of a measurement of $\qs$ in the state $\rho_t$ (note also [41]). Quantum uncertainties in the reaction yields were investigated \cite{PRA2017} in the context of precision limits of radical-pair magnetometers, but with no relevance to quantum coherence. We will here unravel the connection of the uncertainty $\delta Y$ in the singlet reaction products with the coherent dynamics of radical-pairs.
\section{Wigner-Yanase connection between coherence of reactants and fluctuations of reaction products}
Quantum coherence is about superposition states in a certain basis. Superpositions in the eigenbasis of some observable imply uncertainty in the measurement of this observable. For a simple demonstration, consider a two-dimensional system in a pure state and an observable $A$ having two orthonormal eigenstates $\ket{1}$ and $\ket{2}$ with corresponding eigenvalues $a_1$ and $a_2$. If the system's state $\ket{\psi}$ exhibits coherence in the eigenbasis of $A$, it will be $\ket{\psi}=c_1\ket{1}+c_2\ket{2}$. In this state it is $\braket{A^n}=|c_1|^2a_1^n+|c_2|^2a_2^n$, thus the uncertainty $(\Delta A)_{\ket{\psi}}=\big[\braket{A^2}-\braket{A}^2\big]^{1/2}=(|a_1-a_2|/2){\cal C}_1\llbracket\ket{\psi}\rrbracket$, where ${\cal C}_1\llbracket\ket{\psi}\rrbracket=2|c_1c_2|$ is the $l_1$-coherence measure of $\ket{\psi}$ \cite{PlenioCoh}. Thus, in this simple case zero uncertainty implies zero coherence, and vice versa. Parenthetically, one might argue that this discussion is basis-dependent. Coherence is indeed basis-dependent, but the basis is not always up to us to choose. For example, in spin-chemical reactions studied herein, the singlet-triplet basis is defined by the molecular electronic structure, hence it is not an abstract mathematical choice (see \cite{Yu} for a further example involving the controlled-NOT gate).
\begin{figure}[t]
\begin{center}
\includegraphics[width=8 cm]{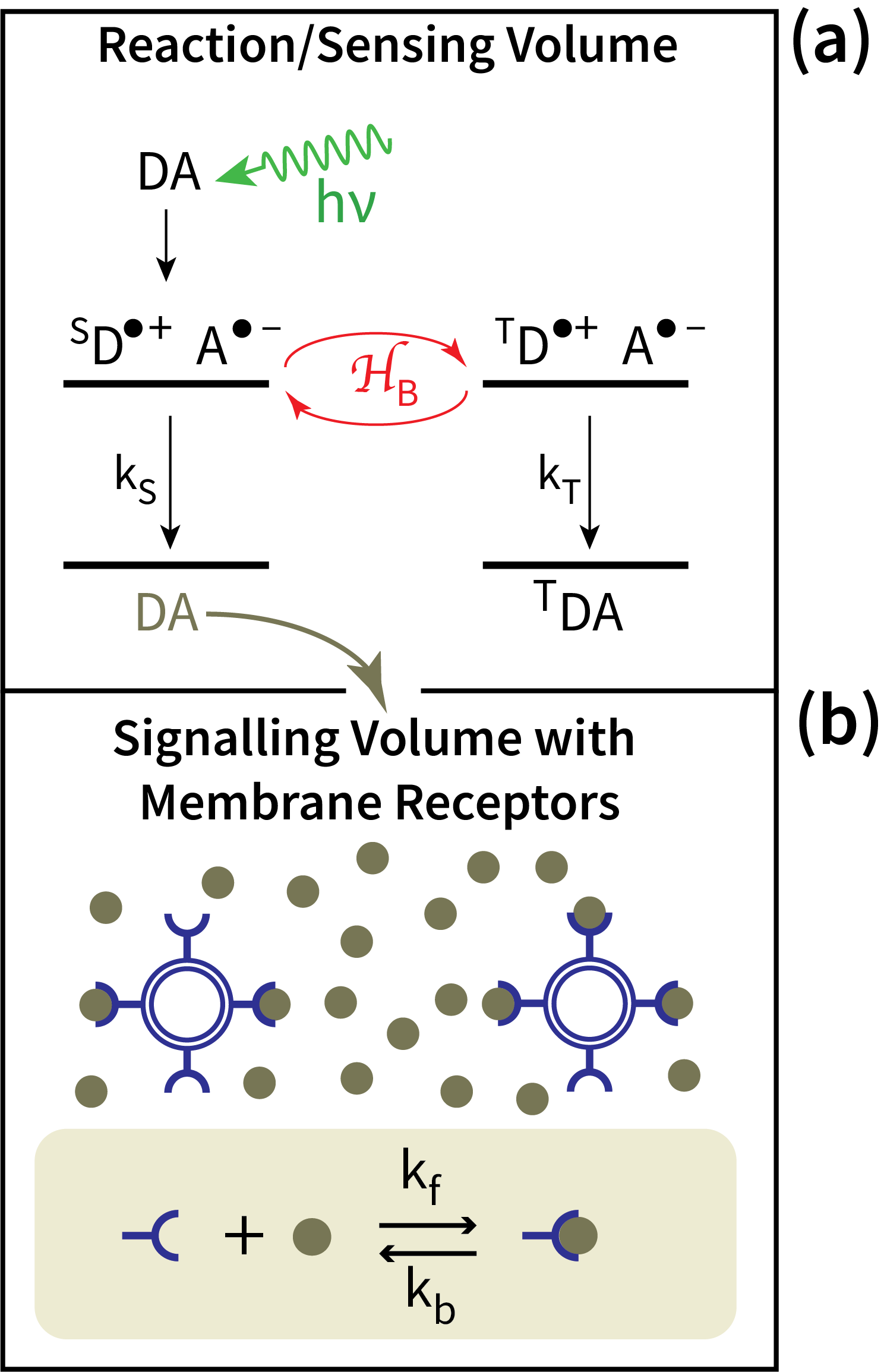}
\caption{Radical-pair reaction scheme and physiological transduction mechanism based on membrane receptors. (a) Schematic of radical-pair reactions, showing the light-induced charge separated state, the singlet radical-pair $^{\rm S}{\rm D}^{\bullet +}{\rm A}^{\bullet -}$ undergoing coherent mixing with the triplet radical-pair $^{\rm T}{\rm D}^{\bullet +}{\rm A}^{\bullet -}$, driven by the hamiltonian ${\cal H}_B$. Radical-pairs in any coherent superposition of singlet and triplet states can recombine to either the singlet or the triplet channel, the reaction yields being magnetic-field-dependent. (b) A cellular receptor-ligand system conveys the information on the changing magnetic field through the changing number of  receptor-ligand complexes. Such complexes form when the ligands, the singlet reaction products shown as blue circles, bind to membrane receptors in the "sensing" volume. Shown in the inset is the receptor-ligand binding-dissociation reaction with the relevant rates. This cellular model was introduced in \cite{Weaver}.}
\label{fig1}
\end{center}
\end{figure}
Radical-pair spin states are in general mixed and have dimension larger than 2, hence we need to move beyond the previous simple example. We now demonstrate the connection between coherence and uncertainty for a general mixed state $\rho$ of a system having dimension $d$. Central to this discussion is the Wigner-Yanase \cite{WY} skew information, extensively used in exploring quantum resource theories \cite{Luo1,Luo2,Luo3,Luo4,Yu,Luo5,Luo6}. It is $I_{\rm WY}(\rho,A)=-{1\over 2}\tr\{[\sqrt{\rho},A]^2\}$, where $A$ is an observable in the space spanned by the basis vectors $\ket{j}$, with $j=1,2,..,d$. Yu introduced \cite{Yu} the coherence quantifier
\beq
{\cal C}\llbracket\rho\rrbracket=\sum_{j=1}^dI_{\rm WY}(\rho,\Pi_j),\label{Crho}
\eeq
where $\Pi_j=\ket{j}\bra{j}$ is the projector to the basis state $\ket{j}$. If $A$ has as eigenbasis the states $\ket{j}$ with corresponding eigenvalues $a_j$, here ordered as $a_1<a_2<...<a_d$,  and the system state is $\rho=\sum_{ij}\rho_{ij}\ket{i}\bra{j}$, then the variance of $A$ in this state is $(\Delta A)_\rho^2={1\over 2}\sum_{i\neq j}\rho_{ii}\rho_{jj}(a_i-a_j)^2\geq {1\over 2}(a_1-a_2)^2\sum_{i\neq j}\rho_{ii}\rho_{jj}$. It is similarly found that $\sum_j(\Delta\Pi_j)_\rho^2=\sum_{i\neq j}\rho_{ii}\rho_{jj}$. To make further progress, we will use a formal inequality \cite{Luo1}, namely for any operator $B$, the Wigner-Yanase information $I_{\rm WY}(\rho,B)$ is bounded above by the variance, $(\Delta B)_\rho^2$, of the operator $B$ in the state $\rho$:
\beq
I_{\rm WY}(\rho,B)\leq (\Delta B)_\rho^2\label{Iineq}
\eeq 
Using the definition \eqref{Crho} and the previous expressions for $(\Delta A)_\rho^2$ and $\sum_j(\Delta\Pi_j)_\rho^2$, we thus we arrive at the inequality 
\beq
(a_1-a_2)^2{\cal C}\llbracket\rho\rrbracket\leq 2(\Delta A)_\rho^2 
\eeq
In words, the coherence of the state $\rho$ expanded in the eigenbasis of $A$ is bounded above by the twice the variance of $A$ in that state, given by $(\Delta A)_\rho^2=\tr\{\rho A^2\}-\tr\{\rho A\}^2$. It is noted that in the discussions \cite{Luo6} connecting coherence with uncertainty, one can theoretically distinguish the genuine quantum part of the uncertainty resulting from measurements in a general mixed state $\rho$, from the classical noise inherent in $\rho$. We will not make this distinction, as the mixed states under consideration herein will lead to measurement uncertainties that cannot  be physically split in classical versus quantum, this is why by "uncertainty" we refer to the total uncertainty of an observable as defined previously. 

As noted in the introduction, the singlet-triplet basis is central in radical-pairs, since radical-pair reaction yields are determined by a quantum measurement in the S-T basis. Thus, in order to explicitly address S-T coherence, we need to revise the definition of Eq. \eqref{Crho}, which pertains to coherence within the total basis of the system. Like we did in \cite{PRR2020} by using the relative entropy between $\rho$ and its diagonal version in the S-T space, $\qs\rho\qs+\qt\rho\qt$, we here need to quantify coherence between the whole singlet and the whole triplet subspace of the radical-pair. To that end, we introduce the Wigner-Yanase S-T coherence measure 
\beq
{\cal C}_{\rm ST}\llbracket\rho\rrbracket=I_{\rm WY}(\rho,\qs)+I_{\rm WY}(\rho,\qt)\label{Cst}
\eeq
This is a direct application of the coherence measure introduced by Luo and Sun \cite{Luo4}, which reads ${\cal C}\llbracket\rho\rrbracket=\sum_{i=1}^mI_{\rm WY}(\rho,M_i)$, where $M_i\geq 0$, and $\sum_{i=1}^m\sqrt{M_i}\sqrt{M_i}=\sum_{i=1}^m M_i=\mathsf{1}$ define a general quantum measurement. Here we set $m=2$, with $M_1=\qs$, $M_2=\qt$. Using again the inequality \eqref{Iineq}, and that the variance of $\qs$ equals the variance of $\qt$ (see derivation of Eq. \eqref{dY}), we obtain 
\beq
{\cal C}_{\rm ST}\llbracket\rho\rrbracket\leq 2(\Delta\qs)_\rho^2\label{CSTQS}
\eeq
Like in \cite{PRR2020,Kattnig}, we define a single number quantifying S-T coherence over the whole reaction, the reaction-averaged coherence
\beq
{\cal C}_{\rm ST}=\int_0^\infty{\cal C}_{\rm ST}\llbracket\tilde{\rho}_t\rrbracket ke^{-kt}dt, 
\eeq
because again, the coherence, ${\cal C}_{\rm ST}\llbracket\tilde{\rho}_t\rrbracket$, of the time-dependent radical-pair spin state, $\tilde{\rho}_t$, must be weighted by the reaction term $ke^{-kt}$. Using the expression \eqref{dY} for the reaction yield uncertainty $\delta Y$ together with \eqref{CSTQS} we obtain our first main result, {\it the reaction-integrated Wigner-Yanase S-T coherence is bounded above by two times the variance of the singlet reaction yield}:
\beq
{\cal C}_{\rm ST}\leq 2(\delta Y)^2\label{bound}
\eeq
The implication of this inequality can be hardly overstated. {\it It connects an S-T coherence measure of the reactants with physiologically measurable fluctuations in the number of reaction products}. Such relation could have wider applicability for quantum biological effects beyond the particular paradigm of radical-pair reactions discussed here. We also note that the bound \eqref{bound} is formal, i.e. valid for all radical-pairs of any dimension (any number of nuclear spins), and any kind of  hamiltonian interactions or relaxation processes. The formal bound \eqref{bound} resulted from the introduction of the S-T coherence measure \eqref{Cst} based on the Wigner-Yanase information and the formal bound \eqref{Iineq} connecting the Wigner-Yanase information with the variance of the associated operator.

Admittedly, the smaller the upper bound in \eqref{bound}, the better this inequality confirms the {\it absence} of coherence. The converse in obviously not true, i.e. a large upper bound only implies the {\it possibility} of a large coherence, and thus the merit of a deeper analysis of the biological process under consideration. 

Next, reaction yield uncertainties will be connected to the statistics of physiological observables related to cellular processing of the radical-pair reaction products. Based on the connection \eqref{bound} between S-T coherence and uncertainty, we will then define a physiological search for quantum-coherent effects. To address the dynamic cellular processing of radical-pair reaction products with e.g. the singlet product concentration changing with time around an operating value, Weaver and coworkers \cite{Weaver} developed the receptor-ligand model shown in Fig. \ref{fig1}b. A flux of ligand molecules (the reactants DA) enters the "reaction volume", and a fraction $Y$ of this flux (the singlet reaction products DA) enters the "sensing volume". The sensing volume contains {\it in total} $R$ membrane receptors, which can bind the ligand molecules. As receptor-ligand binding is ubiquitous in biology, the basic idea of this model is that the number $C$ of receptor-ligand bound complexes will further signal at the physiological level the magnetic-field changes reflected in the changing concentration of the ligands (reaction products).
\section{Coherence is a resource in a receptor-ligand physiological signaling system}
Receptor-ligand complexes are formed at the first-order rate $k_f$ and broken at the zeroth-order rate $k_b$, i.e. it will be $dC/dt=k_f(R-C)L-k_bC$, where $R-C$ is the number of unoccupied receptors, and $L\propto Y$ is the concentration of the ligands, proportional to the singlet reaction yield. In the steady-state it will be $dC/dt=0$, from which follows 
\beq
C={{KR}\over {L+K}},\label{Ceq}
\eeq
where $K=k_b/k_f$ is the equilibrium constant of the ligand-receptor binding reaction. Since $L\propto Y$, a magnetic field change $\delta B$ will effect a change in ligand concentration by 
\beq
\Delta L={L\over Y}\Big|{{dY}\over {dB}}\Big|\delta B,\label{deltaL}
\eeq
and thus a change in $C$ derived from \eqref{Ceq} 
\beq
\Delta C={{KR}\over {(L+K)^2}}\Delta L\label{deltaC} 
\eeq
As noted in \cite{Weaver}, inherent in the ligand-receptor system are fluctuations in $C$ given by 
\beq
\delta C={\sqrt{LKR}\over{L+K}}\label{dc}
\eeq
These are classical Poisson fluctuations stemming from the probabilistic nature of the receptor-ligand binding at the single molecule level \cite{Lauffenburger}. Finally, the authors in \cite{Weaver} identify with $\Delta C$ the biochemical "signal" conveying the magnetic field change, and with $\delta C$ the relevant "noise". Requiring $\Delta C\geq \delta C$ they arrive at the minimum number of receptors necessary to achieve a desired magnetic sensitivity $\delta B_{\rm min}$, i.e. $R\geq {{(L+K)^2}\over {KL}}\big({Y\over {|dY/dB|\delta B_{\rm min}}}\big)^2$, which is minimized at $L=K$, the minimum being
\beq
R_{\rm min}=\Big({{2Y}\over {|dY/dB|\delta B_{\rm min}}}\Big)^2\label{RtW}
\eeq
The rationale of the minimization is \cite{Weaver} that since $R$ is a biological resource, evolutionary pressure should have led to a system design where the number of receptors $R$ is minimum, given the required $\delta B_{\rm min}$.

However, besides considering the classical noise $\delta C$ of Eq. \eqref{dc}, we can now introduce in the discussion the reaction yield quantum uncertainty $\delta Y$ of Eq. \eqref{dY}. The corresponding uncertainty in the ligand concentration $L$ is
\beq
\widetilde{\delta L}=\sqrt{L\over V}\delta Y,\label{dLq}
\eeq
where $V$ is the sensing volume. 

We can now look at the previous discussion of \cite{Weaver} from a different perspective, building on the quantum foundations of the system under study. Firstly, and before moving to the level of the receptor-ligand transducing system, we can conclude that the magnetic field change $\delta B$ is measurable {\it by whatever means} only if the change in ligand concentration, $\Delta L$, produced by the change $\delta B$ is larger than the quantum noise $\widetilde{\delta L}$. Requiring $\Delta L\geq \widetilde{\delta L}$ and using \eqref{deltaL}, \eqref{dLq} leads to
\beq
\delta B\geq{1\over \sqrt{LV}}{{Y\delta Y}\over {|dY/dB|}}\label{dB}
\eeq
Thus we recover the standard quantum limit to the magnetic sensitivity \cite{Budker}, scaling as $1/\sqrt{LV}=1/\sqrt{\rm number~of~ligands}$, which demonstrates the consistency of the previous calculations.

Moving to the next layer, the ligand-receptor transducing system, and in light of Eq. \eqref{deltaC} relating $\Delta C$ with $\Delta L$, the requirement $\Delta L\geq \widetilde{\delta L}$ is equivalent to $\Delta C\geq \widetilde{\delta C}$, where $\widetilde{\delta C}$ is the quantum noise in $C$ stemming from $\widetilde{\delta L}$. That is, to find $\widetilde{\delta C}$ we replace $\Delta L$ with $\widetilde{\delta L}$ in Eq. \eqref{deltaC}. The relation $\Delta C\geq \widetilde{\delta C}$ does not provide any more information beyond the bound \eqref{dB}. It just restates the fundamental measurability of $\delta B$ in the receptor-ligand transducing layer. The biochemical signal conveying a magnetic-field-change is still $\Delta C$, but now the fundamental noise limiting measurability of $\delta B$ is not the classical noise $\delta C$ but the quantum noise $\widetilde{\delta C}$.

Now, however, we can propose another design rationale for determining the biological resource $R$. We can here argue that nature could determine $R$ by the requirement that the receptor/ligand classical noise $\delta C$ of Eq. \eqref{dc} {\it does not add extra noise beyond $\widetilde{\delta C}$,  the quantum fluctuations in} $C$. In other words, the requirement to have a \enquote{quantum-limited} receptor/ligand transducing system, well-adapted to the quantum fluctuations in the reaction yield. This assumption is physically allowed, nevertheless at this point it is just an assumption. Moreover, it is a realistic assumption, due to the analogy with the human visual system's perception limits set by the photon statistics of the stimulus light, as outlined in Sec. II. Here we unravel the consequences of such an assumption. We show that precise measurements of physiological observables and their scaling behaviour can reflect quantum coherent effects.

The aforementioned requirement translates to $\delta C\leq \widetilde{\delta C}$, which leads to 
\beq
R\geq {{(L+K)^2V}\over K}{1\over {(\delta Y)^2}}\label{RtK}
\eeq
This is apparently minimized for $L=0$, but then $\delta B$ shoots up according to \eqref{dB}. Thus we minimize the expression $R(\delta B)^2$, which by \eqref{dB} and \eqref{RtK} reads $R(\delta B)^2\geq {{(L+K)^2}\over {KL}}\big({Y\over {|dY/dB|}}\big)^2$. Alas, we arrive again at the same value for the combined quantity $R_{\rm min}(\delta B_{\rm min})^2$, as found previously in Eq. \eqref{RtW} along the lines of Weaver and co-workers. Summarizing, by bringing into the discussion the quantum noise $\widetilde{\delta C}$ we introduced two requirements, $\Delta C\geq\widetilde{\delta C}$ leading to \eqref{dB}, and $\widetilde{\delta C}\geq \delta C$ leading to \eqref{RtK}. The former inequality reflects the fundamental measurability of $\delta B$, while the latter expresses the assumption of a quantum-limited design. Together, these two inequalities imply $\Delta C\geq\delta C$, which resulted from the previous rationale that led to \eqref{RtW} along the lines of \cite{Weaver}. Thus the rationale of \cite{Weaver} still holds, now being an implication of a "quantum" design rationale of the receptor-ligand system.

Only now we obtain additional insights, since by applying the minimization condition $L=K$ we arrive at two expressions, one for the sensing figure of merit $\delta B_{\rm min}$ following from \eqref{dB}, and one for the biological resource $R_{\rm min}$ following from \eqref{RtK}:
\begin{align}
(\delta B_{\rm min})^2&={1\over {KV}}\Big({{Y\delta Y}\over {|dY/dB|}}\Big)^2\label{dbRtmin1}\\
R_{\rm min}&={{4KV}\over {(\delta Y)^2}}\label{dbRtmin2}
\end{align}
We can now use our first main result, the bound \eqref{bound} connecting the variance of the reaction product with the coherence of the reactants. Combining \eqref{bound} with \eqref{dbRtmin2} we find
\beq
R_{\rm min}\leq {{8KV}\over {{\cal C}_{\rm ST}}}\label{Rtbound}\\
\eeq
The inequality \eqref{Rtbound} is our second main result, connecting an S-T coherence measure of the radical-pairs (reactants) with measurable physiological quantities of the first biochemical layer transducing the reaction products, like the sensing volume $V$, the reaction constant $K$ of the transducing system, and the number of receptors $R_{\rm min}$. {\it Large coherence is seen to imply the necessity for fewer receptors, showing that quantum coherence is indeed a biological resource that can propagate in biochemical layers outside the coherent core process}.
\section{Singlet-triplet coherence as a constraint for a biological resource and a biological figure of merit}
Finally, it would be useful to formulate a similar connection of the magnetic sensitivity $\delta B_{\rm min}$ with the reaction-averaged coherence ${\cal C}_{\rm ST}$. This is less straightforward to obtain, since  the quantity $Y\delta Y/|dY/dB|$ entering \eqref{dbRtmin1} exhibits a non-trivial dependence on ${\cal C}_{\rm ST}$. Hence we resort to a numerical simulation (see following subsection). The empirical result found from this simulation is that $\delta B_{\rm min}$ is bounded below by 
\beq
\delta B_{\rm min}\gtrapprox{0.015\over {\sqrt{KV}}}{1\over \sqrt{{\cal C}_{\rm ST}}}\label{dBbound}
\eeq
Again, it appears that S-T coherence is a resource for magnetic sensing, since the larger the coherence, the smaller $\delta B_{\rm min}$. The scaling $\delta B_{\rm min}\propto 1/\sqrt{KV}$ is amenable to experimental verification. If observed, the underlying coherence ${\cal C}_{\rm ST}$ can be extracted from \eqref{dBbound}.
\begin{figure*}[th!]
\begin{center}
\includegraphics[width=17.5 cm]{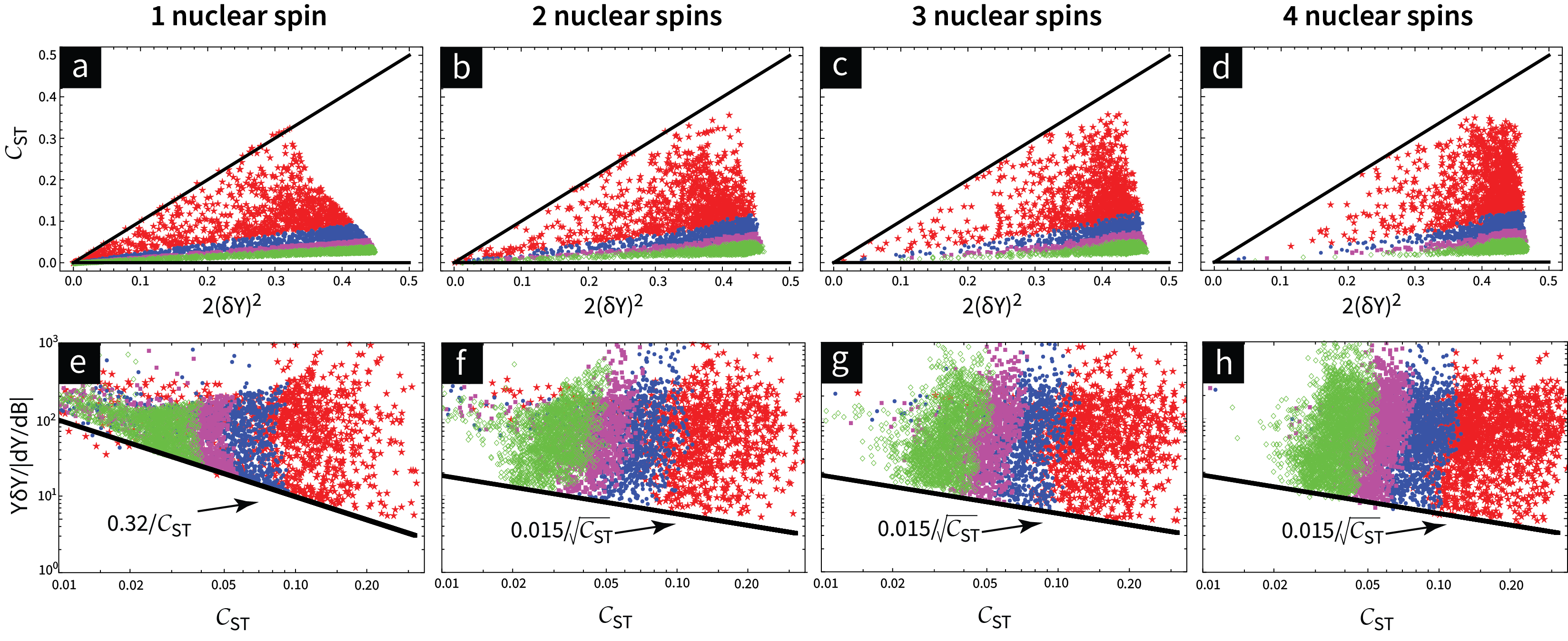}
\caption{Quantum simulation of the Wigner-Yanase singlet-triplet coherence bound and the magnetic sensitivity of radical-pair reactions. The simulation includes 5000 runs for each radical-pair model with number of nuclear spins ranging from 1 to 4. (a-d) The bound ${\cal C}_{\rm ST}\leq 2(\delta Y)^2$ is obviously satisfied, as it is formally derived and holds for any radical-pair model. Points closer to saturating the bound, ${\cal C}_{\rm ST}=2(\delta Y)^2$, correspond to low values of $\kappa_{ST}$. Note that the maximum value of ${\cal C}_{\rm ST}$ is 1/2, while the maximum value of $(\delta Y)^2$ is 1/4. The radical-pair state is always mixed and exhibits partial coherence, hence these maxima are never reached. Large uncertainty does not imply large coherence, since highly mixed states in the S-T basis are uncertain yet incoherent. Note also that some points appear to slightly violate the bound. This is because of the finite time resolution of the simulation, i.e. we split time from $t_{min}=0$ to $t_{max}=10/k$ into 10000 intervals. (e-h) The magnetic sensitivity as given by the ratio $Y\delta Y/|dY/dB|$ entering Eq. \eqref{dbRtmin1}. It is seen that singlet-triplet coherence is a resource, since the magnetic sensitivity is bounded below by $1/{\cal C}_{\rm ST}^m$, with $m=1$ for the minimal radical-pair model with one nuclear spin and $m=1/2$ for more realistic radical-pair models having two or more nuclear spins.
}
\end{center}
\label{fig2}
\end{figure*}
Further assuming that the inequalities \eqref{Rtbound} and \eqref{dBbound} are roughly saturated, we obtain a relation constraining the product of $R_{\rm min}$ (the biological resource, ideally small) with $(\delta B_{\rm min})^2$ (the biological figure of merit, ideally small), by the Wigner-Yanase coherence:
\beq
R_{\rm min}(\delta B_{\rm min})^2\approx{0.1\over {\cal C}_{\rm ST}^{3/2}}\label{unc}
\eeq
{\it With increasing coherence, the system might choose to reduce the required resources or enhance its figure of merit (reduce $\delta B_{\rm min}$)}, {\it however not arbitrarily, but satisfying the constraint \eqref{unc}, which can be seen as a quantum-biological uncertainty relation}. 

Clearly, relation \eqref{unc} does not reflect a necessity, but a possibility. In other words, we here unravel what is physically possible. Whether nature has opted to operate in the parameter regime saturating the bounds \eqref{Rtbound} and \eqref{dBbound}, and thus satisfying the relation \eqref{unc}, remains to be discovered. In any case, the previous considerations outline a specific methodology to search for quantum coherence effects in a layer beyond the quantum-coherent core process, in close proximity to the physiological environment. 
\subsection{Quantum simulation with a multi-spin radical-pair}
To find the relation between $\delta B_{\rm min}$ and ${\cal C}_{\rm ST}$ we perform a full quantum simulation by propagating the master equation $d\rho_t/dt=-i[{\cal H}_B,\rho_t]-\kappa_{ST}(\qs\rho_t+\rho_t\qs-2\qs\rho_t\qs)-k \rho$, with $k=1$ and the hamiltonian ${\cal H}_B={\cal H}_{hf}+B(s_{1z}+s_{2z})+J\mathbf{s_1}\cdot\mathbf{s}_2$, which includes the hyperfine coupling hamiltonian ${\cal H}_{hf}$, the Zeeman interaction of the donor's and acceptor's electron spin, considering the magnetic field along the z axis, and the exchange coupling of the two electronic spins. We have performed the simulation starting with the "minimal' radical-pair model having just one nuclear spin (density matrix is $8\times 8$), which is frequently used in theoretical considerations \cite{HorePRL}, as it is often sufficient to convey the basic physics of the problem under consideration. Nevertheless, we also performed the simulation with more realistic radical-pair models having up to 4 spin-1/2 nuclei, for which case the density matrix is $64\times 64$.

For the minimal radical-pair model the hyperfine hamiltonian is ${\cal H}_{hf}=a\mathbf{s}_1\cdot\mathbf{I}$, where $\mathbf{I}$ is the single nuclear spin operator, here coupled to the donor's electron spin. For a radical-pair model with two or more nuclear spins we also randomize their location, i.e. whether they reside at the donor or at the acceptor. For example, for two nuclear spins, in the former case it would be ${\cal H}_{hf}=a_1\mathbf{s}_1\cdot\mathbf{I}_1+a_2\mathbf{s}_1\cdot\mathbf{I}_2$, whereas in the latter case it would be ${\cal H}_{hf}=a_1\mathbf{s}_1\cdot\mathbf{I}_1+a_2\mathbf{s}_2\cdot\mathbf{I}_2$. We proceed similarly with radical-pair models containing 3 or 4 spin-1/2 nuclei. All couplings and rates are normalized to $k=1$. We randomize the hyperfine couplings $0\leq a_j\leq 10$, the exchange coupling $|J|\leq 10$, and the S-T dephasing rate $0\leq\kappa_{ST}\leq 5$. The initial state is $\rho_0=\qs/\tr\{\qs\}$, i.e. the two electrons in the singlet state and the nucleus fully mixed. 

First, in Figs. 3a-d we numerically verify the bound \eqref{bound}. As noted before, the bound \eqref{bound} is formal, hence the fact that it is satisfied in the numerical simulation is no surprise. What might be less expected is the result shown in Figs. \ref{fig2}e-h, where we plot $Y\delta Y/|dY/dB|$ as a function of ${\cal C}_{\rm ST}$. The derivative $|dY/dB|$ is calculated from the difference $|Y(B=k)-Y(B=0)|/k$, in other words the difference of the singlet reaction yield at the non-zero magnetic field $B=k$ from the yield at $B=0$, divided by the non-zero magnetic field. We thus quantify the low-magnetic-field effect \cite{lowB1}, which makes radical-pair reactions work as magnetometers at close to zero magnetic field. For the minimal radical-pair model having just one nuclear spin it is evident from Fig. \ref{fig2}e that $Y\delta Y/|dY/dB|\gtrapprox 0.32/{\cal C}_{\rm ST}$, whereas for two or more nuclear spins we observe in Figs. \ref{fig2}f-h that $Y\delta Y/|dY/dB|\gtrapprox 0.015/\sqrt{{\cal C}_{\rm ST}}$. Assuming this remains the case for radical-pairs with an even larger number of nuclear spins, we arrive at the bound \eqref{dBbound}. We note that in expressions like $\delta B_{\rm min}\gtrapprox{0.015\over {\sqrt{KV}}}{1\over \sqrt{{\cal C}_{\rm ST}}}$, the magnetic field has units of frequency, since in the simulation of Fig. \ref{fig2} we have normalized all rates, including the magnetic field, to the recombination rate taken as $k=1$. Using the gyromagnetic ratio once can translate frequency into magnetic field.
\section{Conclusions}
To put our findings into perspective, we have presented a connection of the Wigner-Yanase information with coherence and uncertainty in a biological context. Such connection was possible because of the mathematically established bound relating the Wigner-Yanase information with the variance of the relevant operator, and our introduction of a singlet-triplet coherence quantifier derived from the Wigner-Yanase information. 

From the biological side this connection was based on the radical-pair biochemical magnetometer. Similar considerations would apply to estimating the field's direction in the context of the radical-pair biochemical compass. It is also seen that the underlying spin-dependent reaction fades away towards the end of our discussion. What remains from the actual system is just the Wigner-Yanase coherence quantifier connected with the system's figure of merit (the magnetic sensitivity) and a system's basic biological resource (the number of membrane receptors binding to reaction products, the concentration of which is magnetic-field dependent). Our approach unravels physically possible effects and connections in the context of quantum biology working at the physiological level, and can thus drive a systematic search for other quantum-biological effects, since "quantum coherent core process $\rightarrow$ ligand concentration $\rightarrow$ receptors" could be a pervasive biological paradigm.

\end{document}